# Mid-to-far infrared tunable perfect absorption
# by a sub - λ/100 nanofilm in a fractal phasor resonant cavity

Johann Toudert,*,[1] Rosalia Serna,*,[1] Marina García Pardo,[1] Nicolas Ramos,[1]
Ramón J. Peláez,[2] and Belén Maté[2]

[1] *Laser Processing Group, Instituto de Óptica, IO, CSIC, Madrid, Spain*
[2] *Instituto de Estructura de la Materia, IEM, CSIC, Madrid, Spain*
*Corresponding authors: johann.toudert@gmail.com, rosalia.serna@csic.es*

Integrating an absorbing thin film into a resonant cavity is the most practical way to achieve perfect absorption of light at a selected wavelength in the mid-to-far infrared, as required to target blackbody radiation or molecular fingerprints. The cavity is designed to resonate and enable perfect absorption in the film at the chosen wavelength λ. However, in current state-of-the-art designs, a still large absorbing film thickness (~ λ/50) is needed and tuning the perfect absorption wavelength over a broad range requires changing the cavity materials. Here, we introduce a new resonant cavity concept to achieve perfect absorption of infrared light in much thinner and thus really nanoscale films, with a broad wavelength tunability using a single set of cavity materials. It requires a nanofilm with giant refractive index and small extinction coefficient (found in emerging semi-metals, semi-conductors and topological insulators) backed by a transparent spacer and a metal mirror. The nanofilm acts both as absorber and multiple reflector for the internal cavity waves, which after escaping follow a fractal phasor trajectory. This enables a totally destructive optical interference for a nanofilm thickness more than 2 orders of magnitude smaller than λ. With this remarkable effect, we demonstrate angle-insensitive perfect absorption in sub - λ/100 bismuth nanofilms, at a wavelength tunable from 3 to 20 μm.

Perfect absorbers tuned to absorb light at a selected wavelength in the mid-to-far infrared (typically, from 3 μm to 20 μm) are crucial for applications in the biomedical, environment, or security areas.[1] They are needed to target blackbody radiation or molecular fingerprints, as required for thermal recognition,[1] hyperspectral imaging,[2] stealth and cloaking,[3] thermal radiation spectrum shaping.[4] Besides allowing spectral selectivity, perfect absorbers must present a nanoscale thickness. When a perfect absorber acts as active medium in a photodetector, it must be nanoscale-thick to ease photocarrier extraction.[5] In general in any device, nanoscale-thick absorbers are also beneficial for the performance since it is easier to grow defect-free absorbing materials with good physico-chemical properties as thin layers than as thick ones. Another requirement for a perfect absorber to be fully efficient is angle-insensitivity: it must effectively absorb the light impinging from any angle of incidence.

The most advanced kind of nanoscale-thick infrared perfect absorbers is based on metasurfaces.[2-4, 6-13] Metasurfaces show a broadly tunable optical response, however it is often intrinsically angle-dependent. Furthermore, their applicability is limited by their complex fabrication process combining deposition, lithography, etching, which is unpractical and costly. For this reason, efforts have been made for developing thin films with valuable optical properties by more accessible lithography-free deposition processes.[5, 14-27] To achieve nanoscale perfect absorption at a selected wavelength, the most practical approach consists in integrating an absorbing thin film into a simple resonant cavity. Perfect absorption in this film is achieved by destructive optical interference at the cavity resonance wavelength, which can be tuned by changing the component materials and/or tailoring the cavity structure, for instance the absorbing film thickness. The current state-of-the-art resonant cavity design, in terms of structure simplicity and absorbing film thickness needed to achieve perfect absorption at a given wavelength, consists of a strongly absorbing film (refractive index n ~ extinction coefficient k) backed by a finite optical conductivity mirror (finite n < k). The cavity resonance (and thus perfect absorption) is achieved for film thicknesses t down to 1/60 of the wavelength (i.e. $\lambda/60$).[19, 21] This enables a much more compact design than for a standard Fabry-Pérot cavity, for which perfect absorption is achieved for t ~ $\lambda/16$ using standard semiconductor absorbers (e.g. Si, Ge, InAs, InSb, PbS, HgCdTe). This state-of-the-art cavity design has enabled angle-insensitive perfect absorption at selected infrared wavelengths between 5 µm and 12 µm, t being adjusted from 100 nm to 600 nm.[22-25] In other words, still a quite thick absorbing film is needed, especially if perfect absorption in the far infrared is targeted. Furthermore, finite infrared conductivity cannot be achieved with standard metals, which behave as near-perfect conductors at infrared wavelengths because of their high charge carrier density. As a solution, other materials, sometimes artificial (e.g. sapphire, AZO, heavily doped Si) are used, however none presents a finite optical conductivity over the whole infrared spectrum. Therefore, bringing perfect absorption to markedly different infrared wavelengths requires changing the nature of the mirror. Summarizing, with this current state-of-the-art cavity design, achieving infrared perfect absorption requires a still quite thick absorbing film and tuning the perfect absorption wavelength over the mid-to-far infrared is a challenging task.

Here, to achieve perfect absorption in much thinner, really nanoscale films (with a thickness as small as $\lambda/200$) with a facile wavelength tuning over the mid-to-far infrared, a new resonant cavity concept is introduced and applied. It is based on a nanofilm with giant refractive index and small extinction coefficient, backed by a transparent spacer and a near-perfectly conducting mirror. The key element of such cavity is the nanofilm, which acts both as absorber and multiple reflector for the internal cavity waves over a broad wavelength range. After escaping the cavity, these waves follow a fractal phasor trajectory that drives their interference with the wave directly reflected at the air/nanofilm interface. This enables angle-insensitive perfect absorption in films of spectacularly small thickness. A material with the required properties for the nanofilm can be found among the semi-metals, semiconductors and topological insulators of the p-block. The mirror and spacer can consist of any standard metal and low index transparent material, respectively. As a remarkable example, we demonstrate both theoretically and experimentally that a cavity consisting of a bismuth (Bi) nanofilm backed by an $Al_2O_3$ spacer and an Ag film enables angle-insensitive perfect absorption in a sub - $\lambda/100$ Bi nanofilm, at a wavelength tunable in a broad infrared region from 3 to 20 µm by varying only the nanofilm and spacer thicknesses. For instance, perfect absorption can be achieved at $\lambda = 4$ µm and 18 µm in a nanofilm with thickness of only 30 nm ($\sim \lambda/130$) and 90 nm ($\sim \lambda/200$), respectively. In sum, the new kind of cavity we introduce (called hereafter "fractal phasor resonant cavity") enables perfect absorption in much thinner films than with the current state-of-the-art cavities. Also, in contrast with them, there is no need to change the nature of the cavity materials to tune the perfect absorption wavelength over the whole mid-to-far infrared.

Prior to focusing on such novel fractal phasor resonant cavity, we explore the advantages of building a "boosted" Fabry-Pérot cavity based on an absorbing nanofilm with a giant refractive index n and small extinction coefficient k, a solution which has not been considered in the past. As an example, in Figure 1a it is shown the design of such a cavity where the nanofilm with t = 100 nm and a giant n = 10 is backed by a near-perfectly conducting mirror. Figure 1b shows the reflectance spectrum of this cavity, assuming k = 0.5. A resonance with zero reflectance, i.e. perfect absorption, occurs near $\lambda = 5$ µm, i.e. for t ~ $\lambda/50$ and with 80% of the power absorbed in the film (20% lost to the mirror). As seen in Figure 1c, the wavelength of this resonance is independent of k, and it becomes weaker (non-perfect absorption) when k increases. This behavior is typical of a Fabry-Pérot interference mechanism, as well as the cavity signature in the phasor diagram (Figure 1d).

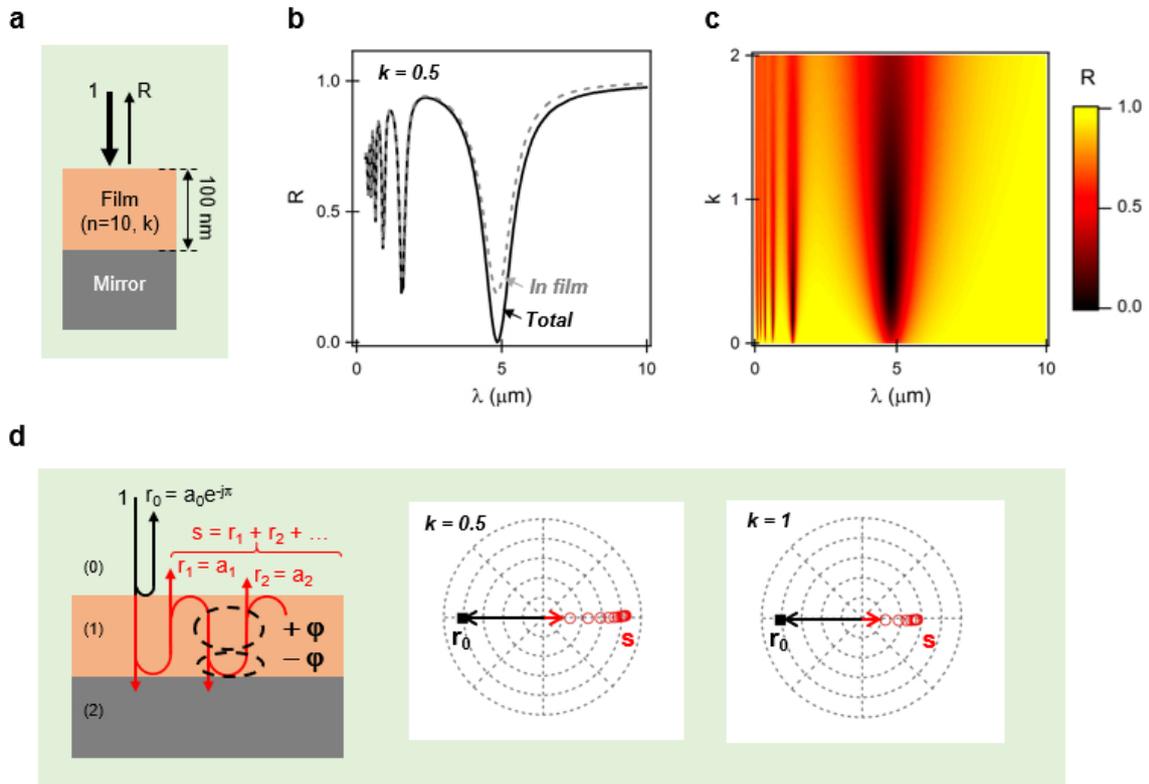

**Fig. 1. "Boosted" Fabry-Pérot resonant cavity for infrared perfect absorption in a λ/50 nanofilm: optical properties and perfect absorption mechanism.** (a) Simplified representation of the cavity. It consists of a nanofilm with giant refractive index n, small extinction coefficient k and thickness t backed by a near-perfectly conducting mirror. Here, as example, we take n = 10, t = 100 nm, and the mirror is made of Ag (with n and k from ref. 28). (b) Simulated reflectance spectrum of the cavity at normal incidence for k = 0.5 (black line) showing perfect absorption at the resonance wavelength λ ∼ 5 μm (t ∼ λ/50), and reflectance loss due to absorption in the film (dashed grey line). (c) Color map of the simulated reflectance spectra for different values of k, showing that perfect absorption is not achieved for larger k. (d) Simplified representation of the wave propagation in the cavity and phasor diagrams at the resonance wavelength λ ∼ 5 μm, for k = 0.5 and k = 1. The vector $r_0$ accounts for the wave directly reflected at the air/film interface (first reflection) and s is the sum of the vectors accounting for the internal cavity waves escaping after multiple reflections ($r_j$'s). These vectors follow a straight line, typical of a Fabry-Pérot cavity. For k = 0.5, s cancels with $r_0$, this totally destructive interference enables perfect absorption.

This phasor diagram describes the interference between the wave directly reflected at the air/nanofilm interface (first reflection, $r_0$) and the internal cavity waves escaping after multiple reflections ($r_1$, $r_2$, etc…) at the resonance wavelength (λ ∼ 5 μm). The vector accounting for $r_0$ points toward the left, while the vectors accounting for the $r_j$'s point toward the right. For k = 0.5, the sum (s) of such vectors is large enough to cancel with $r_0$ (totally destructive interference, i.e. perfect absorption). In contrast, for k = 1, the amplitude of the $r_j$'s is too small to enable a total cancelation (partially destructive interference, i.e. non-perfect absorption). Further details about the perfect absorption mechanism in such kind of cavity are given in Supporting Information S1. Because of the Fabry-Pérot type of such mechanism, the perfect absorption wavelength can be tuned by varying the film thickness. A variation of t up to 500 nm enables tuning the perfect absorption wavelength in the whole 3-20 μm range (Supporting Information S2). Much larger thicknesses are needed if using as absorber a standard semiconductor with lower refractive index (Supporting Information S3).

In sum, this "boosted" Fabry-Pérot cavity rivals the current state-of-the-art cavities in terms of simplicity and film thickness needed to achieve perfect absorption at a given wavelength. It also makes much more practical the tuning of the perfect absorption wavelength over the whole mid-to-far infrared, because only controlling the film thickness is required with no need of changing the nature of the cavity materials. Yet, a drawback of this cavity is that the extinction coefficient of the absorbing film must be small enough to enable perfect absorption. This condition is difficult to fulfil with existing materials in a broad wavelength range. However, this restriction is overcome by an improved design that introduces a transparent spacer between the absorbing film and the mirror,[5] i.e. with the fractal phasor resonant cavity. This design is described on an example in Figure 2a. In this example, a nanofilm with t = 40 nm, n = 10 and k = 1 is backed by a transparent spacer and a near-perfectly conducting mirror. Such nanofilm thickness is sufficient to achieve perfect absorption at λ ∼ 5 μm (Figure 2b), the same as with a 100 nm – nanofilm with lower k in a "boosted" Fabry-Pérot cavity. Therefore, the fractal phasor resonant cavity not only enables to succesfully overcome the restriction on the k value: it also allows to strongly reduce the absorbing film thickness needed to achieve perfect absorption at a given wavelength. Furthermore, near 100% of the power is absorbed in the film (Figure 2b), in contrast with the "boosted" Fabry-Pérot cavity shown in Figure 1 where a significant amount of power is lost to the mirror. In sum, with the fractal phasor resonant cavity, perfect absorption occurs *in* a sub - $\lambda/100$ nanofilm (in this example, t $\sim \lambda/125$), i.e. much thinner than with the current state-of-the-art cavity designs. This spectacular result is allowed by the special optical interference mechanism of the cavity, which is described in the phasor diagram shown in Figure 2c.

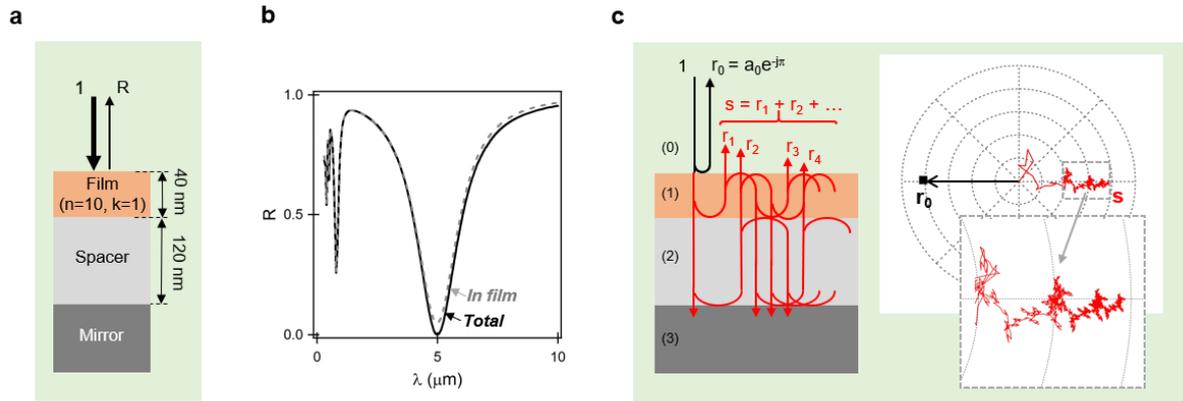

**Fig. 2. Fractal phasor resonant cavity for infrared perfect absorption in a sub - $\lambda/100$ nanofilm: optical properties and perfect absorption mechanism.** (a) Simplified representation of this cavity. It consists of a nanofilm with giant refractive index n, small extinction coefficient k and thickness t, backed by a transparent spacer and a near-perfectly conducting mirror. Here, as example, we take n = 10, k = 1, t = 40 nm, an $Al_2O_3$ spacer (with n = 1.65, k = 0), and an Ag mirror. (b) Simulated reflectance spectrum of the cavity at normal incidence (black line) showing perfect absorption at the resonance wavelength $\lambda \sim 5$ µm (t $\sim \lambda/125$), and reflectance loss due to absorption in the film (dashed grey line). At the resonance wavelength, near 100% of the incident power is absorbed in the film. (c) Simplified representation of the wave propagation in the cavity and phasor diagram at the resonance wavelength $\lambda \sim 5$ µm. The vector $r_0$ accounts for the first reflection and s is the sum of the vectors accounting for the internal cavity waves escaping after multiple reflections ($r_j$'s). These vectors $r_j$'s follow a fractal trajectory. This fractal trajectory yields a s vector that cancels with $r_0$, enabling totally destructive interference, and thus perfect absorption in a particularly thin absorbing film.

As in any resonant cavity, the reflection properties are governed by the interference between the first reflection ($r_0$) and the internal cavity waves escaping after multiple reflections ($r_1$, $r_2$, etc…). However, the giant n and small k of the absorbing nanofilm open a surprising path for these waves inside the cavity before they escape. The nanofilm acts as an efficient multiple reflector for such internal cavity waves (reflections at the nanofilm/air, nanofilm/spacer, spacer/nanofilm interfaces), which can jump many times between the nanofilm and spacer and make many trips in the cavity before escaping, while being weakly absorbed during each trip. In the phasor diagram of the cavity, this translates into a fractal trajectory for the vectors accounting for the $r_j$'s. At the cavity resonance wavelength ($\lambda \sim 5$ μm), the fractal branch has grown fully toward the right and s cancels with $r_0$. Note that off-resonance (Supporting Information S4), such fractal branch is rotated and not fully grown, so that s does not cancel with $r_0$. To the best of our knowledge, this is the first time that such phasor trajectory is reported for resonant cavities.

To bring such finding to the real world, materials fulfilling the above - defined criteria (giant n, small k) must be found. At such aim, one should pay a special attention to semi-metals, semiconductors and topological insulators of the p-block, which are becoming of increasing interest to the photonics and optoelectronic communities.[29] One single-element p-block material, the semi-metal bismuth (Bi), is a particularly interesting candidate as it presents a giant n and small k (8 < n < 10 and 1 < k < 2) over the whole 3 μm to 20 μm wavelength region, as shown in Figure 3a.[29, 30] Other interesting candidates among p-block materials with simple compositions are the semi-metal Sb,[29] the topological insulator $Bi_2Te_3$,[31] and the semiconductor PbTe,[32, 33] which show a higher n than standard semiconductors in the infrared (yet smaller than that of Bi, and with a smaller spectral range of absorption).

To assess the potential of the fractal phasor resonant cavity concept with a nanofilm consisting of Bi, simulations of the optical properties of a $Bi/Al_2O_3/Ag$ cavity (Figure 3b) have been performed. For comparison, simulations done for a Bi/Ag "boosted" Fabry-Pérot cavity are shown in Supporting Information S5. The Bi and $Al_2O_3$ thicknesses, $t_{Bi}$ and $t_{Al2O3}$, have been varied. As shown on the color maps of Figure 3c, which represent the reflectance spectra of the cavity as a function of $t_{Bi}$ for two different $t_{Al2O3}$ values, perfect absorption is achieved at a wavelength that can be tuned over the whole 3 - 20 μm range by varying both thicknesses. As shown in Figure 3d, by choosing suitably these thicknesses, perfect absorption can be achieved with near 100% of the power absorbed in a very thin Bi nanofilm whatever the targeted wavelength. For instance, perfect absorption can be achieved at $\lambda$ = 4 µm and 18 µm in a Bi nanofilm with $t_{Bi}$ = 30 nm ($t_{Bi} \sim \lambda/130$) and 90 nm ($t_{Bi} \sim \lambda/200$), respectively.

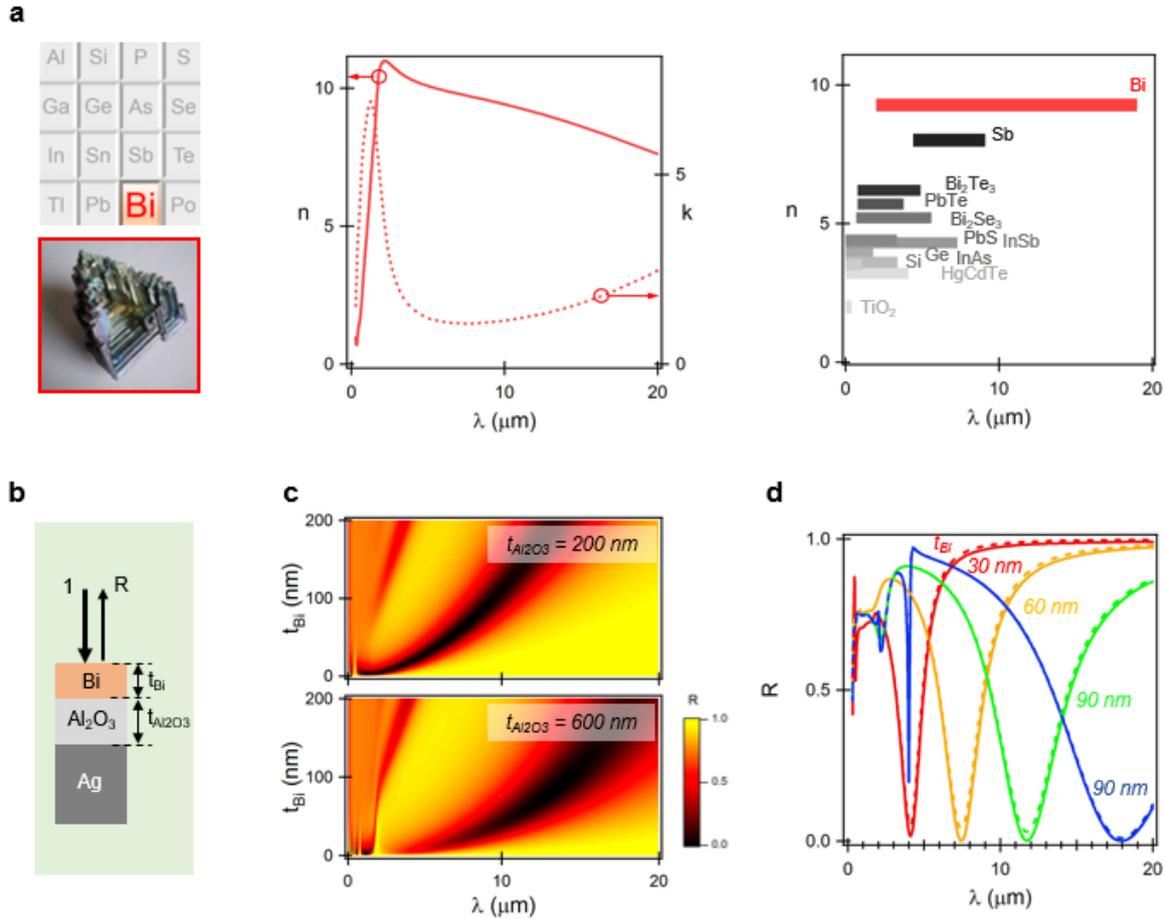

**Fig. 3. Fractal phasor resonant cavity enabling infrared perfect absorption in a sub - $\lambda$/100 Bi nanofilm: tunability of the perfect absorption wavelength in the mid-to-far infrared.** (a) Left panel: Bi is a natural single-element material of the p-block of the periodic table. Middle panel: it shows a giant refractive index (8 < n < 10) and a small extinction coefficient (1 < k < 2) in the whole mid-to-far infrared (3 to 20 µm). Right panel: Some semi-metals, semi-conductors and topological insulators of the p-block present very high n values, higher than those of the "standard" infrared materials (HgCdTe, InAs, InSb, PbS). The horizontal bands represent, for each material, the sub-bandgap spectral region where n takes high values, and the average of such values. (b) Simplified representation of the considered fractal phasor resonant cavity, with a Bi/Al$_2$O$_3$/Ag structure. (c) Color map showing the simulated reflectance spectra of this cavity at normal incidence as a function of the Bi film thickness $t_{Bi}$ and Al$_2$O$_3$ transparent spacer thickness $t_{Al2O3}$. The spacer adds a degree of freedom that enables shifting the perfect absorption wavelength in the whole the 3 to 20 µm region with small $t_{Bi}$ values. (d) Simulated reflectance spectra of this cavity for selected ($t_{Bi}$, $t_{Al2O3}$) values (color lines), and corresponding reflectance loss due to absorption in the Bi film (color dashed lines). Perfect absorption is reached for all the spectra with near 100% absorption in the Bi film. The ($t_{Bi}$, $t_{Al2O3}$) values of the different spectra are (30 nm, 100 nm), (60 nm, 200 nm), (90 nm, 400 nm) and (90 nm, 1200 nm), where $t_{Bi}$ represents the following fractions of the resonance wavelength: $\sim \lambda$/130, $\lambda$/150, $\lambda$/130, $\lambda$/200.

To demonstrate such properties in an experiment, Bi/Al$_2$O$_3$/Ag cavities have been fabricated by depositing Bi films on Al$_2$O$_3$/Ag stacks grown on Si. As a reference, Bi/Ag Fabry-Pérot cavities have also been grown on Si. The materials have been grown by physical deposition, following procedures detailed elsewhere.[30, 34] The aspect of the Bi target used for deposition and some of the fabricated samples are shown in Figure 4a (upper panel). The cross-section images of the grown Bi/Al$_2$O$_3$/Ag cavities (Figure 4a, lower panel) show a well-defined layered organization, with the continuous Bi film on top. Bi thicknesses less than 100 nm are observed.

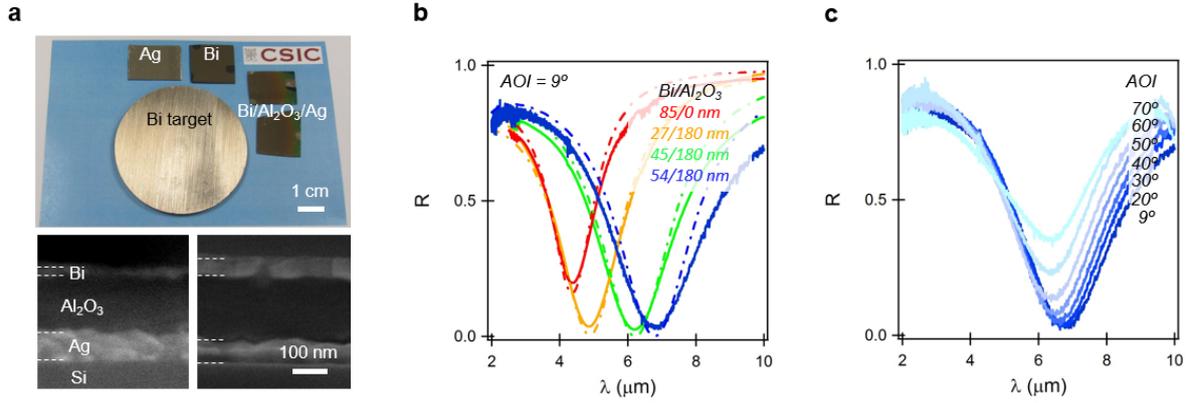

**Fig. 4. Fractal phasor resonant cavity enabling infrared perfect absorption in a sub - λ/100 Bi nanofilm: experimental demonstration of the perfect absorption wavelength tuning and angle-insensitivity.** (a) top panel: picture of the Bi target used to grow the nanofilms by physical deposition, and of some of the fabricated cavity samples. Bottom panels: cross-section images of some of the fabricated fractal phasor resonant cavities (Bi/Al$_2$O$_3$/Ag cavity). These images confirm the layered structure, and the small Bi film thicknesses, of few tens of nm. (b) Reflectance spectra of the cavities with different Bi thicknesses, at near normal incidence (9º). Continuous lines represent the experimental data and dash-dotted lines represent the corresponding simulations. Perfect absorption is achieved at a wavelength λ tuned from 4.8 to 6.8 μm by varying t$_{Bi}$ from 27 to 54 nm (~ λ/180 to λ/120). The spectrum of a Bi/Ag Fabry-Pérot cavity is also shown. For this cavity, perfect absorption is not achieved. (c) Reflectance spectra of the Bi/Al$_2$O$_3$/Ag cavity with t$_{Bi}$ = 54 nm as a function of the angle of incidence. Near perfect absorption remains in a wide angular range, up to 45º.

Figure 4b shows the reflectance spectra of the cavities, at near-normal incidence (9º). The cavities resonate at different wavelengths in the mid-infrared that depend on the type of cavity (Bi/Al$_2$O$_3$/Ag or Bi/Ag) and Bi thickness. The Bi/Ag cavity with t$_{Bi}$ = 85 nm resonates at λ = 4.2 μm (t$_{Bi}$ ~ λ/50). However, it is far to enable perfect absorption, due to the too large extinction coefficient k of Bi. In contrast, as expected, perfect absorption is nearly achieved with the Bi/Al$_2$O$_3$/Ag cavities. For Bi films with t$_{Bi}$ = 27 nm, 45 nm, and 54 nm, the cavity resonates at λ = 4.8 μm (t$_{Bi}$ ~ λ/180), 6.2 μm (t$_{Bi}$ ~ λ/140) and 6.8 μm (t$_{Bi}$ ~ λ/125), respectively. In addition to their spectral selectivity, the Bi/Al$_2$O$_3$/Ag cavities show angle-independent optical properties. This is exemplified in the reflectance spectra of Figure 4c, which show that the perfect absorption is maintained for angles of incidence up to 45°. The observed trend is well reproduced by simulations (Supporting Information S6).

In conclusion, the theoretical and experimental results shown in this work demonstrate that the new fractal phasor resonant cavity concept we propose enables overcoming the current state-of-the-art cavity solutions in terms of absorber compactness and tuning robustness. In the examples, we show the successful achievement of infrared perfect absorption in films with much smaller thicknesses (t down to $\lambda/200$, versus ∼ $\lambda/50$ for current state-of-the-art designs) and there is no need to change the nature of the cavity materials to tune the perfect absorption wavelength over the whole mid-to-far infrared: only the spacer and nanofilm thickness must be adjusted. In addition, the resonance wavelength is not sensitive to the nature of the mirror and the spacer (Supporting Information S7), making the use of cheaper or more convenient materials than Ag and Al$_2$O$_3$ possible. Note that, although we chose to use the semi-metal Bi as absorbing material because of its champion optical properties, other materials (semi-metals, semiconductors, or topological insulators) from the p-block could be used. Currently, there is a growing interest in fabricating p-block compounds with tunable optical properties.[31, 35-38] This quest may unveil new materials with an infrared refractive index even higher than that of Bi, enabling therefore perfect absorption in films with smaller thickness. Note that very high infrared refractive indexes can also be found at the vicinity of the phonon bands of many dielectrics.[39] With a fractal phasor resonant cavity, this effect could therefore enable perfect absorption in very thin films, however in a narrow spectral region. Besides these material aspects, perfect absorption in smaller dimensions may also be achieved with cavity designs exploring different fractal phasor trajectories. Therefore, the new fractal phasor resonant cavity concept we propose may open a pathway toward the facile lithography-free fabrication of few-nm infrared perfect absorbers with broad spectral tunability. In particular, if built from semi-metals or topological insulators, such ultrathin absorbers may enable an extremely strong coupling between infrared light and surface electronic states, ideal for optoelectronic interfacing in the considered biomedical, environment, and security applications.

**Methods**

The reflectance, power absorbed in the film, and phasor calculations were done within the transfer matrix formalism using the WVASE32 software (Woollam Co. Inc.) and a home-made code. The Bi/Ag and Bi/Al$_2$O$_3$/Ag cavities were grown by physical deposition techniques, especially pulsed laser deposition. The layer nominal thicknesses were 100 nm for Ag, 180 nm for Al$_2$O$_3$, and that of Bi was 80 nm for the Bi/Ag structure and it was varied in the range of a few tens of nm for the Bi/Al$_2$O$_3$/Ag structure. The thickness of the deposited Al$_2$O$_3$ was confirmed by ultraviolet-visible-near infrared ellipsometry measurements. The thicknesses of deposited Bi were extracted by fitting the measured infrared reflectance spectra. To realize such fitting, the Bi dielectric function in the infrared was taken from ref. 30, where it was determined on Bi films grown with the same deposition technique. The infrared dielectric function of Al$_2$O$_3$ was determined by extrapolating the measured ultraviolet-visible-near infrared data. The infrared dielectric function of Ag was taken from ref. 28. The so-determined deposited Bi and Al$_2$O$_3$ thicknesses were in good agreement with those observed in the cross-section images obtained with a scanning electron microscope. The infrared reflectance spectra were measured at room temperature with an IFS 66 Bruker Fourier transform infrared spectrometer equipped with a DTGS detector. The incident beam, with a 5 mm diameter, was unpolarized and the angle of incidence was varied from 9º to 70º.

**Author Contributions**

J.T. proposed the concept and did the optical simulations. R.S. coordinated and supervised the experimental work. M.G.P. and N.R. fabricated the cavities and did the basic material characterization. R.P. and B.M. measured the infrared reflectance. J.T. wrote the paper with the advice of R.S. and the input of the other authors.

**Acknowledgments**

M.G.P acknowledges support from the European Social Fund. This work was partially funded by the Spanish Ministry for Economy and Competitiveness through the project MINECO/FEDER TEC2015-69916-C2-1-R.

## S1. "Boosted" Fabry-Pérot cavity: more details about the perfect absorption mechanism

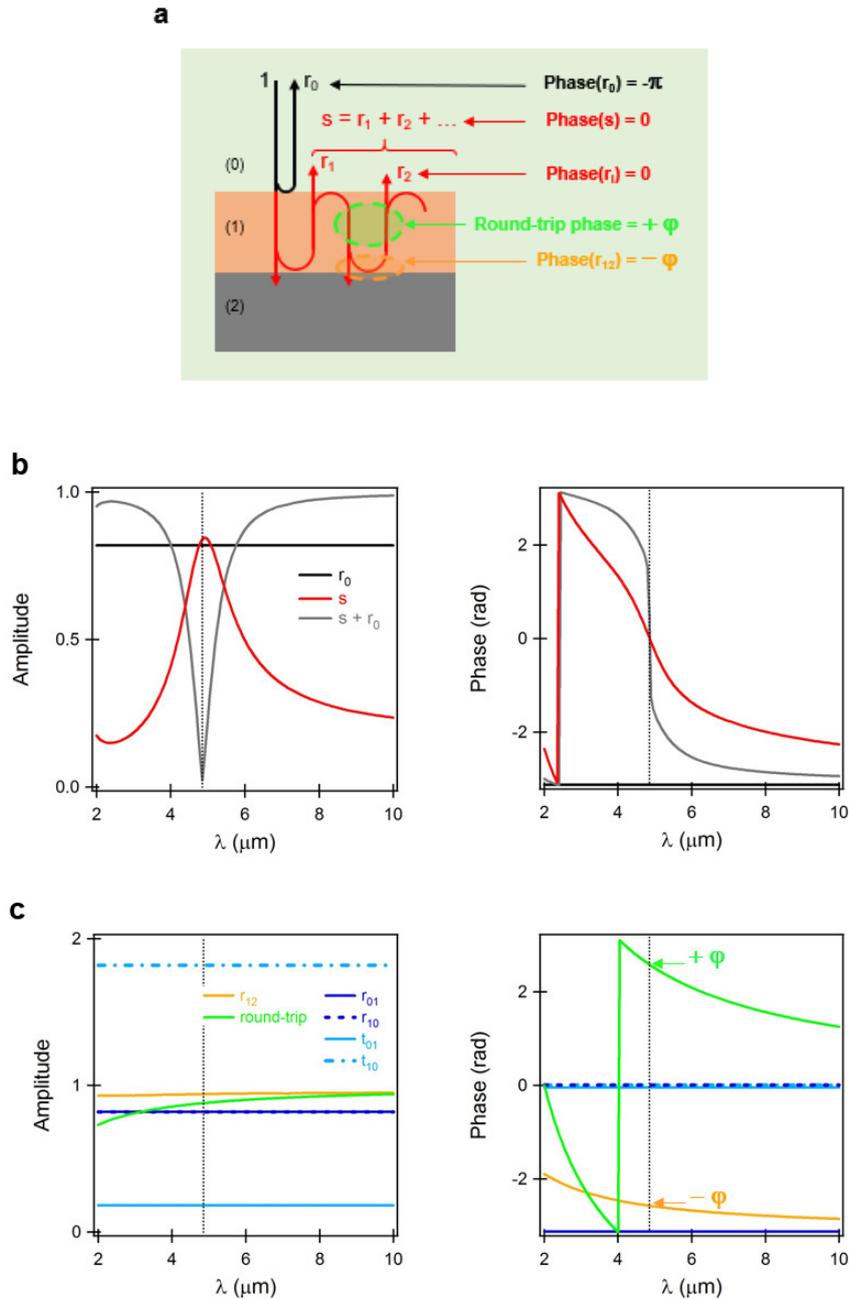

**Fig. S1.** This figure refers to the "boosted" Fabry-Pérot resonant cavity shown in figure 1: film/near-perfectly conducting mirror, with n = 10, k = 0.5, t = 100 nm for the film. (a) Schematic representation of wave propagation in the cavity at the perfect absorption wavelength, at normal incidence. The round-trip phase shift (φ) of an internal cavity wave is exactly opposite to the reflection phase shift (-φ) at the film/mirror interface, 1/2. Therefore, each internal cavity wave escaping the cavity has a null phase shift with respect to the incident wave, and so does their sum (s). Thus, the phase of s is π shifted from that of the wave directly reflected at the air/film interface 0/1 (r0). This enables the totally destructive interference between s and r0. (b) Calculated spectra of the amplitude and phase of r0, s and r0+s, showing the perfect absorption at λ ∼ 5 μm. (c) Calculated spectra of the amplitude and phase of the reflection coefficients at the interfaces between the different media (air = 0, film = 1, mirror = 2), and of the round-trip contribution. All the reflection coefficients have a 0 phase, except $r_{01}$ (which equals to $r_0$) that has a -π phase and $r_{23}$. $r_{23}$ is slightly different from π, because the mirror is non-perfectly conducting. This slightly different phase enables perfect absorption to occur for a film thickness slightly smaller than expected from the Fabry-Pérot formula (λ/50 < λ/4n, i.e. λ/40 with n = 10).

## S2. "Boosted" Fabry-Pérot cavity: tuning of the perfect absorption wavelength in the mid-to-far infrared

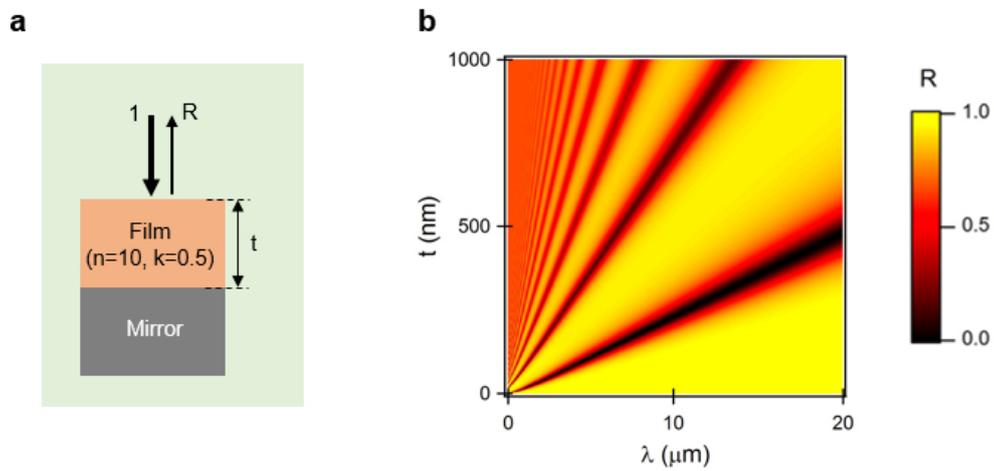

**Fig. S2.** This figure refers to the "boosted" Fabry-Pérot resonant cavity shown in figure 1: film/near-perfectly conducting mirror, with n = 10, k = 0.5, but with an adjustable film thickness t for the film. (a) Simplified representation of the cavity. (b) Color map showing the simulated reflectance spectra of the cavity at normal incidence as a function of t. The perfect absorption wavelength can be tuned in the whole 3 – 20 μm region upon varying the film thickness up to 500 nm.

## S3. "Boosted" Fabry-Pérot cavity vs standard Fabry-Pérot cavity: tuning of the perfect absorption wavelength in the mid-to-far infrared

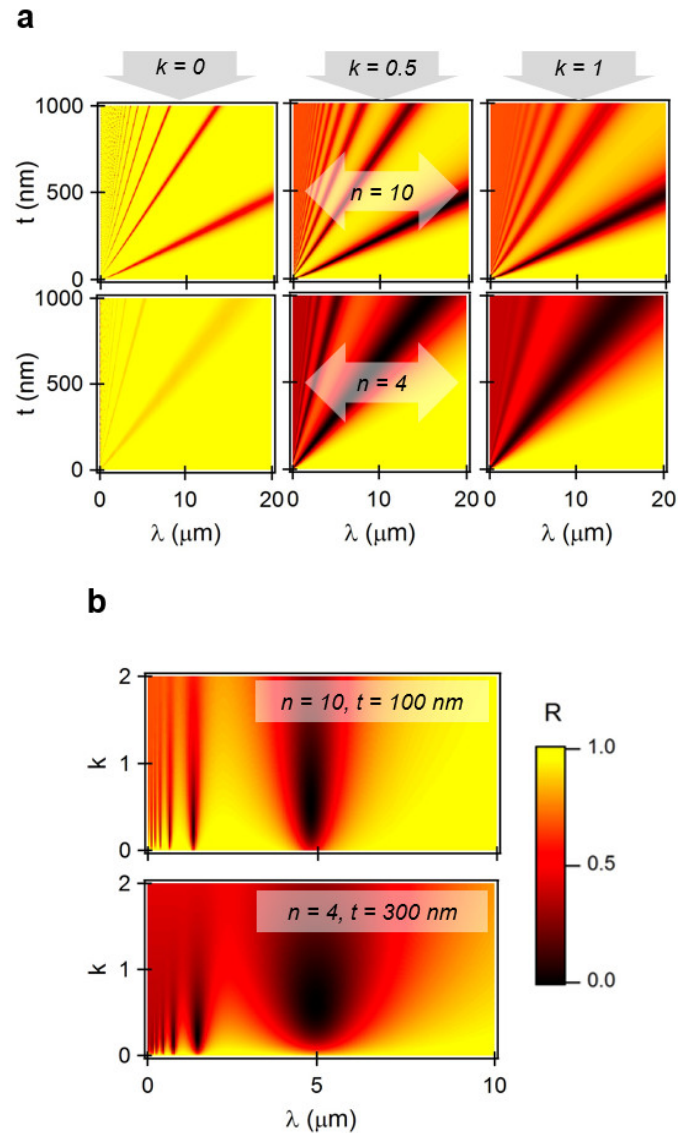

**Fig. S3.** This figure refers to the Fabry-Pérot resonant cavity shown in figure 1: film/near-perfectly conducting mirror, but with adjustable film thickness t, refractive index n, and extinction coefficient k for the film. (a) Color maps showing the simulated reflectance spectra of the cavity at normal incidence as a function of t (from 0 to 1000 nm), n (4 or 10) and k (0, 0.5 or 1). The variation in t needed to tune the resonance wavelength in the whole 3 – 20 µm range decreases if n is larger. Perfect absorption is not achieved if k is too large. (b) Color maps showing the simulated reflectance spectra as a function of k for two cavities with different n and t, but the same resonance wavelength λ ∼ 5 µm. For a lower n, increasing t is necessary to maintain the resonance at the same wavelength, and the resonance is broader.

## S4. Fractal phasor resonant cavity: phasor at different wavelengths

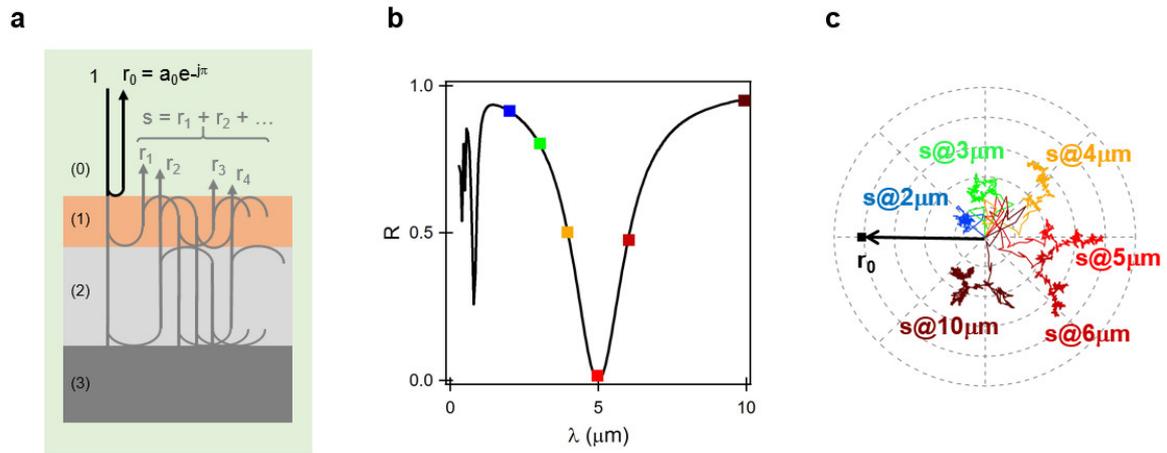

**a**

**b**

**c**

**Fig. S4.** This figure refers to the fractal phasor resonant cavity shown in figure 2: film/transparent spacer/near-perfectly conducting mirror, with n = 10, k =1, t = 40 nm for the film, a 120 nm – thick spacer and an Ag mirror. (a) Schematic representation of wave propagation in the cavity at normal incidence. The vector $r_0$ accounts for the first reflection and s is the sum of the vectors accounting for the internal cavity waves escaping after multiple reflections ($r_j$'s). (b) Simulated reflectance spectrum of the cavity at normal incidence. (c) Phasor diagram of the cavity, at the different wavelengths marked with the same color in (b). The $r_j$'s follow a fractal trajectory that depends on $\lambda$. At the perfect absorption wavelength ($\lambda \sim 5$ µm), the fractal branch is fully grown and yields a s vector that cancels with $r_0$. At non-resonant wavelengths, the fractal branch is rotated and not fully grown, so that s does not cancel with $r_0$.

**S5. Bi/Ag cavity: tuning of the perfect absorption wavelength in the mid-to-far infrared**

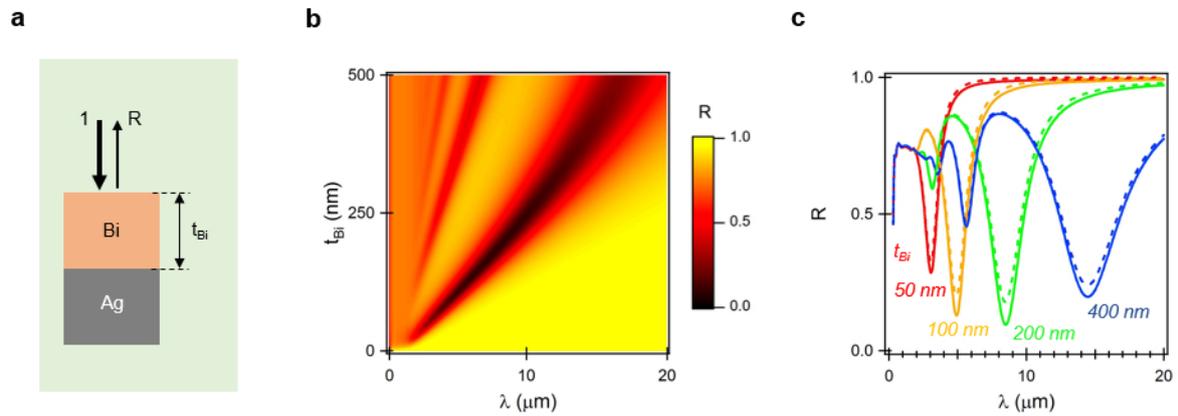

**Fig. S5.** (a) Simplified representation of the Bi/Ag cavity. The n and k of Bi are those shown in figure 3. (b) Color map showing the simulated reflectance spectra of this cavity at normal incidence, as a function of the Bi film thickness $t_{Bi}$. The cavity resonance shifts in the 3 to 18 µm region by varying $t_{Bi}$ up to 500 nm ($t_{Bi} \sim \lambda/40$ to $\lambda/50$) (c) Simulated reflectance spectra of this cavity for selected $t_{Bi}$ values (color lines), and corresponding reflectance loss due to absorption in the Bi film (color dashed lines). Perfect absorption is not achieved with this cavity.

## S6. Bi/Al₂O₃/Ag cavity: Effect of the angle of incidence

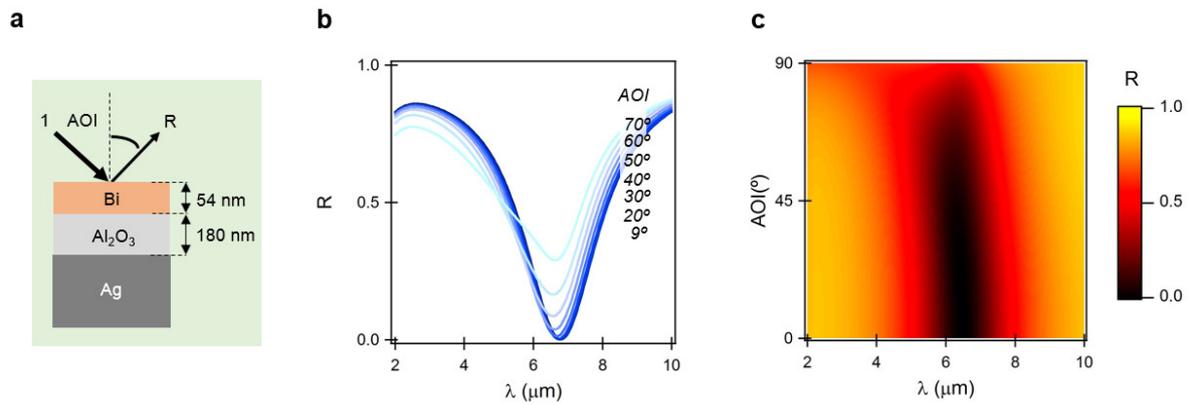

**Fig. S6.** This figure refers to the fractal phasor resonant cavity shown in figure 4: Bi/Al₂O₃/Ag structure with $t_{Bi}$ = 54 nm and $t_{Al2O3}$ = 180 nm. (a) Simplified representation of the cavity. (b) Simulated reflectance spectra of the cavity for different angles of incidence, with unpolarized light. (c) Corresponding color map, showing the reflectance spectra as a function of the angle of incidence. These simulations agree very well with the experimental results in figure 4c.

## S7. Fractal phasor resonant cavity: Material independence

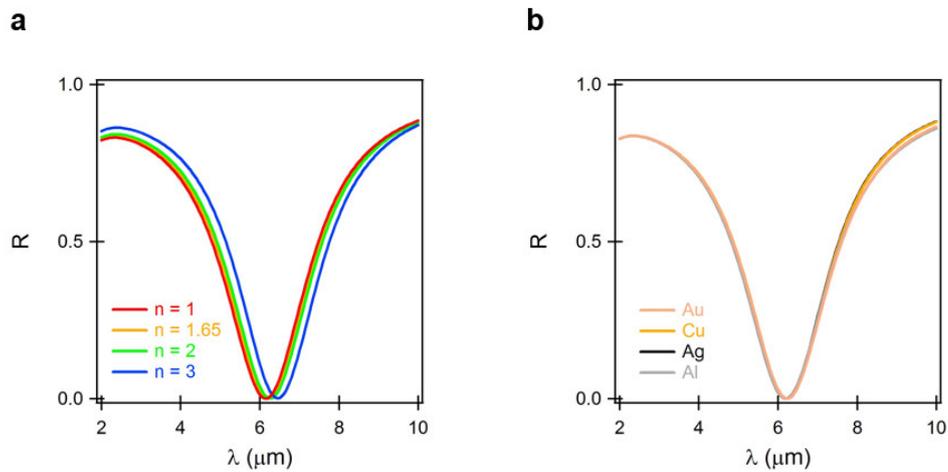

**Fig. S7.** This figure refers to a fractal phasor resonant cavity similar with that shown in figure 4: Bi/transparent spacer/near-perfectly conducting mirror with $t_{Bi}$ = 54 nm and $t_{spacer}$ = 180 nm. Here, the effect of the nature of the spacer (a) and the mirror (b) on the spectral position of the cavity resonance is studied by simulations. No marked shift of the resonance is seen upon varying the refractive index of the spacer from 1 to 3 (a). The resonance is insensitive to the nature of the mirror. Any metal, even the cheap Al or Cu can be used equally.